\documentclass[aps,prd,reprint,amsmath,amssymb,superscriptaddress,floatfix]{revtex4}
\usepackage{graphicx}
\usepackage{verbatim} 
\usepackage{amsfonts}
\usepackage{amssymb}
\usepackage{rotating}
\usepackage{booktabs}
\usepackage{xcolor}
\usepackage{soul}
\usepackage{color}
\usepackage{slashed}
\usepackage{multirow}
\usepackage{makecell}
\usepackage{epsf}
\usepackage{ulem}
\usepackage{cancel}
\usepackage{color,bm}
\usepackage[colorinlistoftodos]{todonotes}

\usepackage[colorlinks=true,citecolor=cyan,urlcolor=blue,bookmarks=true,bookmarks=true,bookmarksopen=true,bookmarksnumbered=true,bookmarksopenlevel=3]{hyperref}

\definecolor{airforceblue}{rgb}{0.36, 0.54, 0.66}
\definecolor{steelblue}{rgb}{0.27, 0.51, 0.71}
\definecolor{amber}{rgb}{1.0, 0.49, 0.0}

\newcommand{\s}{\slashed}

\begin{document}

\title{Single spin asymmetry $A_{UL}^{\sin(2\phi_h-2\phi_R)}$ in dihadron semi-inclusive DIS}
\author{Xuan Luo}
\author{\textsc{Hao Sun}\footnote{Corresponding author: haosun@mail.ustc.edu.cn \hspace{0.2cm} haosun@dlut.edu.cn}}
\affiliation{ Institute of Theoretical Physics, School of Physics, Dalian University of Technology, \\ No.2 Linggong Road, Dalian, Liaoning, 116024, P.R.China }
\date{\today}

\begin{abstract}

The single longitudinal spin asymmetry $A_{UL}^{\sin(2\phi_h-2\phi_R)}$ of dihadron production in semi-inclusive deep inelastic scattering (SIDIS) is examined through helicity-dependent dihadron fragmentation function (DiFF) $G_1^\perp$. The correlation of the longitudinal polarization of a fragmenting quark with the transverse momenta of the produced hadron pair is illustrated by this DiFF. The experimental investigation for this azimuthal asymmetry in dihadron SIDIS by the COMPASS Collaboration has lately yielded a very small signal. Here, the unknown T-odd dihadron fragmentation function $G_1^\perp$ utilizing a spectator model is computed. The model has been successfully used to describe the dihadron production in both the unpolarized and the single polarized processes to access the asymmetry and clarify why the signal is very small. The transverse momentum dependent factorization method, in which the transverse momentum of the final state hadron pair is left unintegrated, has been considered. The $\sin(2\phi_h-2\phi_R)$ asymmetry at the COMPASS kinematics is estimated and we compare it with the data. Besides, the predictions on the same asymmetry are also made at the HERMES and Electron Ion Collider.

\vspace{0.5cm}
\end{abstract}
\maketitle
\setcounter{footnote}{0}

\section{INTRODUCTION}
\label{I}

In hadronization process, there is a nonvanishing probability that at a hard scale a highly virtual parton fragments into two hadrons inside a same jet. This nonperturbative mechanism can be encoded in the so-called dihadron fragmentation functions (DiFFs). The DiFFs were introduced for the first time in Ref.\cite{Konishi:1979cb} and their evolution equations have been investigated in Ref.\cite{Vendramin:1981gi,Vendramin:1981te,Ceccopieri:2007ip}. In particular, the authors of Ref.\cite{Ceccopieri:2007ip} presented the evolution equations 
for extended dihadron fragmentation functions explicitly dependent on the invariant mass, $M_h$, of the hadron pair. Then Ref.\cite{Collins:1994ax} analysed the transversely polarized fragmentation by using the transversely polarized DiFF, which gave rise to the defination of $H_1^\sphericalangle$. The basic physical picture of all possible unpolarized DiFFs was proposed in Ref.\cite{Bianconi:1999cd}. The authors in Ref.\cite{Radici:2001na} expanded the hadron pair system in relative partial waves. By using this approach, some cases that have already been studied in the literatures can now be naturally incorporated in a unified formlism. They also presented new positivity on the DiFFs. Soon after the analysis of DiFFs was extended to the subleading twist whithin a collinear picture \cite{Bacchetta:2003vn}. It is necessary to emphasize that the general expression of the cross section in terms of structure functions for the dihadron SIDIS was proposed whithin transverse momentum dependent (TMD) framework \cite{Gliske:2014wba}. The analysis is complete and up to the subleading twist. Researchers started to keep a watchful eye on the DiFFs when they tried to extract the chiral-odd transversity distribution. The transversity distribution was firstly extracted by considering the Collins effect \cite{Collins:1992kk} in one hadron SIDIS and back-to-back production of dihadron in $e^+ e^-$ annihilations \cite{Anselmino:2008jk}. In this approach, one must apply the TMD factorization framework and consider the QCD evolution of TMDs since two processes under consideration occur at two different scale.
To access the transversity distribution in a more convenient way, an alternative approach considering dihadron SIDIS came to notice which only needs collinear factorization.
Among this mechanism, the chiral-odd DiFF $H_1^\sphericalangle$ \cite{Radici:2001na,Bacchetta:2002ux} couples with $h_1$ at the leading-twist level. The function $H_1^\sphericalangle$ can be extracted from two back-to-back hadron pairs production process in $e^+ e^-$ annihilation \cite{Courtoy:2012ry}. In the literature, the transversity distribution has been extracted from both dihadron SIDIS and proton proton collision data \cite{Bacchetta:2011ip,Bacchetta:2012ty,Radici:2015mwa,Radici:2016lam,Radici:2018iag}.
On the other hand, to estimate the magnitudes of various DIFFs, the model predictions of the DiFFs were performed by the spectator model \cite{Bianconi:1999uc,Bacchetta:2006un,Bacchetta:2008wb,Yang:2019aan,Yang:2019kqw,Luo:2019frz} and by the Nambu-Jona-Lasinio (NJL) quark model \cite{Matevosyan:2013aka,Matevosyan:2013eia,Matevosyan:2017alv,Matevosyan:2017uls}.

Experimentally, the HERMES collaboration \cite{Airapetian:2008sk} produced the experimental data of azimuthal asymmetry in dihadron SIDIS process with a transversely polarized proton target. The COMPASS collaboration \cite{Adolph:2012nw,Adolph:2014fjw} also release the similar experimental data with polarized protons and deuterous targets.
The BELLE collaboration \cite{Vossen:2011fk} have measured the azimuthal asymmetry of a back-to-back two dihadron pair production, reaching the first parameterization of $H_1^\sphericalangle$.
Recently, the COMPASS collaboration \cite{Sirtl:2017rhi} collected the experimental data of various azimuthal asymmetries by scattering longitudinally polarized muons off longitudinally polarized protons.
Therotically, these azimuthal asymmetries appear within the TMD factorization framework, where a $\sin(\phi_h-\phi_R)$ modulation has been studied in the spectator model \cite{Luo:2019frz}. Here $\phi_h$ denotes the azimuthal angle of the hadron pair system and $\phi_R$ is the angle between the lepton plane and two-hadron plane.
In this paper we focus on the $\sin(2\phi_h-2\phi_R)$ modulation.
Within the TMD factorization appoach, the dihadron SIDIS cross section is written as a convolution of transverse momentum dependent parton distribution functions (TMD-PDFs) and TMD-DiFFs. TMD factorization extends collinear factorization by accounting for the parton transverse momentum. In practice, the COMPASS measurement found that the $\sin(2\phi_h-2\phi_R)$ asymmetry is compatible with zero whithin experimental precision.
In this paper, we explore the $\sin(2\phi_h-2\phi_R)$ asymmetry using the spectator model results of the relevant PDFs and DiFFs.
After performing partial waves expansion, the only term contributing to this asymmetry is $g_{1L} G_{1,TT}^\perp$ where $G_{1,TT}^\perp$ origins from interference of two $p$-waves and $g_{1L}$ is the helicity distribution. We adopt the spectator model \cite{Bacchetta:2006un} to calculate $G_{1,TT}^\perp$ and  find that one must consider loop contributions to obtain a nonvanishing $G_{1,TT}^\perp$. Applying the spectator model results for the distributions and DiFFs, we estimate the $\sin(2\phi_h-2\phi_R)$ asymmetry at COMPASS kinematics and compare it with the COMPASS preliminary data.

The paper is organized as follows. In Sec.\ref{II} we review the theoretical framework of the $\sin(2\phi_h-2\phi_R)$ azimuthal asymmetry of dihadron production in unpolarized muon beam scattered off a longitudinally polarized proton target. We apply the spectator model to calculate the T-odd helicity DiFF $G_{1,TT}^\perp$ in Sec.\ref{III}. In Sec.\ref{IV}, we give the numerical results of the $\sin(2\phi_h-2\phi_R)$ azimuthal asymmetry at the kinematics of COMPASS as well as EIC. We summarize our work in Sec.\ref{V}.

\section{The $A_{UL}^{\sin(2\phi_h-2\phi_R)}$ asymmetry in dihadron SIDIS }\label{II}

We consider the SIDIS process of two pions production
\begin{eqnarray} \label{eq1}
\mu(\ell)+p^\rightarrow(P) \to \mu(\ell')+\pi^+(P_1)+\pi^-(P_2)+X,
\end{eqnarray}
where a longitudinally polarized target nucleon possessing a mass $M$, polarization $S$ and momentum $P$, through  the interchange of a virtual photon with momentum $q=\ell-\ell'$, is scattered off by  a unpolarized muon having a momentum $\ell$ . Inside the target, the dynamic quark with momentum $p$ is struck by the photon  and the final state quark with momentum $k=p+q$ then fragments into two leading unpolarized hadrons $\pi^+$ and $\pi^-$ with mass $M_1, M_2$, and momenta $P_1, P_2$. In order to compute the differential cross section pertaining to dihadron-dependent structure function, we express the following kinematic invariants:
\begin{eqnarray} \label{eq2}
\begin{aligned}
x   &=\frac{k^+}{P^+} \qquad  y=\frac{P \cdot q}{P \cdot \ell} \qquad z=\frac{P_h^-}{k^-}=z_1+z_2 \\
z_i &=\frac{P_i^-}{k^-} \qquad  Q^2=-q^2  \qquad s=(P+\ell)^2 \\
P_h &=P_1+P_2 \qquad  R=\frac{P_1-P_2}{2} \qquad  M_h^2=P_h^2  .
\end{aligned}
\end{eqnarray}
The longitudinal light-cone coordinate $\displaystyle a^\pm = \frac{a^0 \pm a^3}{\sqrt{2}}$ and the transverse light-cone coordinate $\vec{a}_T=(a^1,a^2)$ are given in terms of an  arbitrary four vector $a$, in such way  that the component form could be listed as $[a^-,a^+,\vec{a}_T]$.
The light-cone fraction of target momentum taken by the initial quark is designated by $x$, $z_i$ which symbolises the light-cone fraction of hadron $\pi_i$ in terms of the fragmented quark.
The light-cone fraction of fragmenting quark momentum carried by the final hadron pair is  identified by $z$. What's more, the invariant mass, the total momentum and the relative momentum of the hadron pair are represented  by $M_h$, $P_h$ and $R$, respectively.
It is suitable to select the $\hat{z}$ axis consistent with the condition $\vec{P}_{hT}=0$.
Consequently, the momenta $P_h^\mu$, $k^\mu$ and $R^\mu$ can be written as in \cite{Bacchetta:2006un}
\begin{eqnarray} \label{eq3}
\begin{aligned}
P_h^{\mu}&=\left[P^-_h,\frac{M_h^2}{2P^-_h}, \vec{0}_T \right] \\ 
k^{\mu}  &=\left[\frac{P_h^-}{z},\frac{z(k^2+\vec{k}_T^2)}{2P_h^-},\vec{k}_T \right] \\
R^{\mu}  &=\left[-\frac{|\vec{R}|P^-_h}{M_h}\cos\theta,\frac{|\vec{R}|M_h}{2P^-_h}\cos\theta,|\vec{R}|\sin\theta\cos\phi_R,|\vec{R}|\sin\theta\sin\phi_R \right] \\
&=\left[-\frac{|\vec{R}|P^-_h}{M_h}\cos\theta,\frac{|\vec{R}|M_h}{2P^-_h}\cos\theta,  \vec{R}_T^x, \vec{R}_T^y \right] ,
\end{aligned}
\end{eqnarray}
where
\begin{eqnarray} \label{eq4}
\begin{aligned}
|\vec{R}| = \sqrt{\frac{M_h^2}{4}-m_\pi^2}
\end{aligned}
\end{eqnarray}
with $m_\pi$ the mass of pion. It is desired to notice that in order to perform partial-wave expansion, we have reformulated the kinematics in the center of mass frame of the dihadron system. $\theta$ is the center of mass polar angle of the pair with respect to the direction of $P_h$ in the target rest frame \cite{Bacchetta:2002ux}. Here we can find some useful relations as
\begin{eqnarray} \label{eq5}
\begin{aligned}
P_h \cdot R &= 0 \\
P_h \cdot k &= \frac{M_h^2}{2z}+z\frac{k^2+\vec{k}_T^2}{2} \\
R \cdot k &= \left( \frac{M_h}{2z}-z\frac{k^2+\vec{k}_T^2}{2M_h} \right) |\vec{R}| \cos\theta - \vec{k}_T \cdot \vec{R}_T .
\end{aligned}
\end{eqnarray}

The TMD DiFFs $D_1$ and $G_1^\perp$ which will appear in the underlying asymmetry are extracted from the quark-quark correlator $\Delta(k, P_h, R)$
\begin{eqnarray} \label{eq6}
\begin{aligned}
\Delta(k, P_h, R) \displaystyle &=\sum \kern -1.3 em \int_X \;
\int \frac{d^4 \xi}{(2\pi)^4} \; e^{\text{i} k\cdot\xi}\;
\langle 0|\psi(\xi) \,|P_h,R; X\rangle
\langle X; P_h,R|\, \overline{\psi}(0)\,|0\rangle
\Big|_{\xi^- = \vec{\xi}_T  =0} \; \\
&=\frac{1}{16\pi}\left\{ D_1 \slashed{n}_-+G_1^\perp \gamma_5 \frac{\varepsilon_T^{\rho\sigma} R_{T\rho} k_{T\sigma}}{M_h^2} \s{n}_- + \cdots \right\} .
\end{aligned}
\end{eqnarray}
Then we express the leading-twist quark-quark correlator Eq.(\ref{eq6}) in terms of center of mass variables. The connection between the two representations is defined as
\begin{eqnarray} \label{eq7}
\begin{aligned}
\Delta(z,k_T^2,\cos\theta,M_h^2,\phi_R) = \frac{|\vec{R}|}{16zM_h} \int dk^+ \Delta(k, P_h, R) .
\end{aligned}
\end{eqnarray}
By projecting out the usual Dirac structures, we obtain the following decomposition results
\begin{eqnarray} \label{eq8}
\begin{aligned}
4 \pi \text{Tr}[\Delta(z,k_T^2,\cos\theta,M_h^2,\phi_R)\gamma^- \gamma^5] = \frac{\varepsilon_T^{\rho\sigma} R_{T\rho} k_{T\sigma}}{M_h^2}G_{1}^{\perp} ,
\end{aligned}
\end{eqnarray}
where $\gamma^-$ is the negative light-cone Dirac matrix.

The TMD DiFFs $D_1$, $G_1^\perp$ can be expanded in the relative partial waves of the dihadron system up to the $p$-wave level \cite{Bacchetta:2002ux}: 
\begin{eqnarray} \label{eq9}
\begin{aligned}
D_{1}(z,k_T^2, \cos \theta,M_h^2)&=D_{1,OO}+D_{1,OL}\cos \theta+\frac{1}{4}D_{1,LL}(3\cos^2\theta-1)
\\
&+\cos(\phi_k-\phi_R)\sin\theta(D_{1,OT}+D_{1,LT}\cos\theta) + \cos(2\phi_k-2\phi_R)\sin^2\theta D_{1,TT}, \\
G_1^\perp(z,k_T^2, \cos \theta,M_h^2)&=G_{1,OT}^\perp+G_{1,LT}^\perp \cos\theta + \cos(\phi_k-\phi_R) \sin\theta G_{1,TT}^\perp ,
\end{aligned}
\end{eqnarray}
where $G_{1,OT}^\perp$ comes from the interference of $s$- and $p$-waves, and $G_{1,TT}^\perp $ originates from the interference of two $p$-waves with the same transverse polarizations.

Then we will consider azimuthal asymmetries of SIDIS process with unpolarized muons scattering off longitudinally polarized nucleon target. Using TMD factorization approach and denoting $A(y) = 1-y+\frac{y^2}{2}$, the differential cross section for this process reads \cite{Radici:2001na}
\begin{eqnarray} \label{eq10}
\begin{aligned}
\displaystyle \frac{d^9 \sigma_{UU}}{dx dy dz d\phi_S d\phi_h d\phi_R d\cos\theta d\vec{P}_{h\perp}^2 dM_h^2} = \frac{\alpha^2}{2\pi s x y^2} A(y) \sum_q e_q^2 \mathcal{I}[f_1^q D_{1,OO}^q]
\end{aligned}
\end{eqnarray}
and
\begin{eqnarray} \label{eq11} 
\begin{aligned}
&\displaystyle \frac{d^9 \sigma_{UL}}{dx dy dz d\phi_S d\phi_h d\phi_R d\cos\theta d\vec{P}_{h\perp}^2 dM_h^2} 
\\
&=\frac{\alpha^2}{2\pi s x y^2} A(y) \sum_q e_q^2 \sin^2\theta \sin(2\phi_h-2\phi_R) \mathcal{I}\left[ \frac{2(\vec{k}_T \cdot \hat{P}_{h\perp})^2-\vec{k}_T^2}{M_h^2} g_{1L}^q \left( \frac{|\vec{R}|}{2|\vec{k}_T|} G_{1,TT}^\perp \right) \right] , 
\end{aligned}
\end{eqnarray}
where in Eq.(\ref{eq11}) we only resort the term we are interested in, and $\phi_S$ is the azimuthal angles of $\vec{S}_T$ with respect to the lepton scattering plane. 
$\hat{P}_{h\perp}$ satisfies $\hat{P}_{h\perp} =\vec{P}_{h\perp}/|\vec{P}_{h\perp}|$.
For convenience, we have indicated the unpolarized or longitudinally polarized states of the beam or the target with the labels $U$ and $L$, respectively. The structure functions occuring in Eqs.(\ref{eq10}-\ref{eq11}) are written as weighted convolutions of the form
\begin{eqnarray} \label{eq12}
\begin{aligned}
\mathcal{I}[\omega fD] = \int d^2 \vec{p}_T d^2 \vec{k}_T \delta\left( \vec{p}_T-\vec{k}_T-\frac{\vec{P}_{h\perp}}{z} \right)\omega(p_T,k_T)f(x,p_T^2)D(z,k_T^2),
\end{aligned}
\end{eqnarray}
where $\omega(p_T,k_T)$ is an arbitrary function.
In Eq.(\ref{eq10}), $f_1^q$ and $D_{1,OO}^q$ are the unpolarized PDF and unpolarized DiFF with flavor $q$. In Eq.(\ref{eq11}), $g_{1L}^q$ is the helicity distribution function coupled with the T-odd DiFF $G_{1,TT}^{\perp}$. Thus the $\sin(2\phi_h-2\phi_R)$ asymmetry of the considered process can be expressed as
\begin{eqnarray} \label{eq13}
\begin{aligned}
A_{UL}^{\sin(2\phi_h-2\phi_R)}&=&\displaystyle \frac{2}{3} \frac{\sum_q e_q^2 \mathcal{I}\left[ \frac{2(\vec{k}_T \cdot \hat{P}_{h\perp})^2-\vec{k}_T^2}{M_h^2} g_{1L}^q \left( \frac{|\vec{R}|}{2|\vec{k}_T|} G_{1,TT}^\perp \right) \right]}{\sum_q e_q^2 \int \mathcal{I}[f_1^q D_{1,OO}^q]} .
\end{aligned}
\end{eqnarray}

\section{The model calculation of $G_{1,TT}^\perp$}
\label{III}

In this section, we review the model calculation for $G_{1,TT}^\perp$ partly following previous works \cite{Yang:2019aan} and \cite{Luo:2019frz}.
The tree level correlator yields vanishing contributions to $G_{1,TT}^\perp$ on account of shortage of the imarginary phase. 
\begin{figure}[htp]
\centering
\includegraphics[height=3cm,width=4cm]{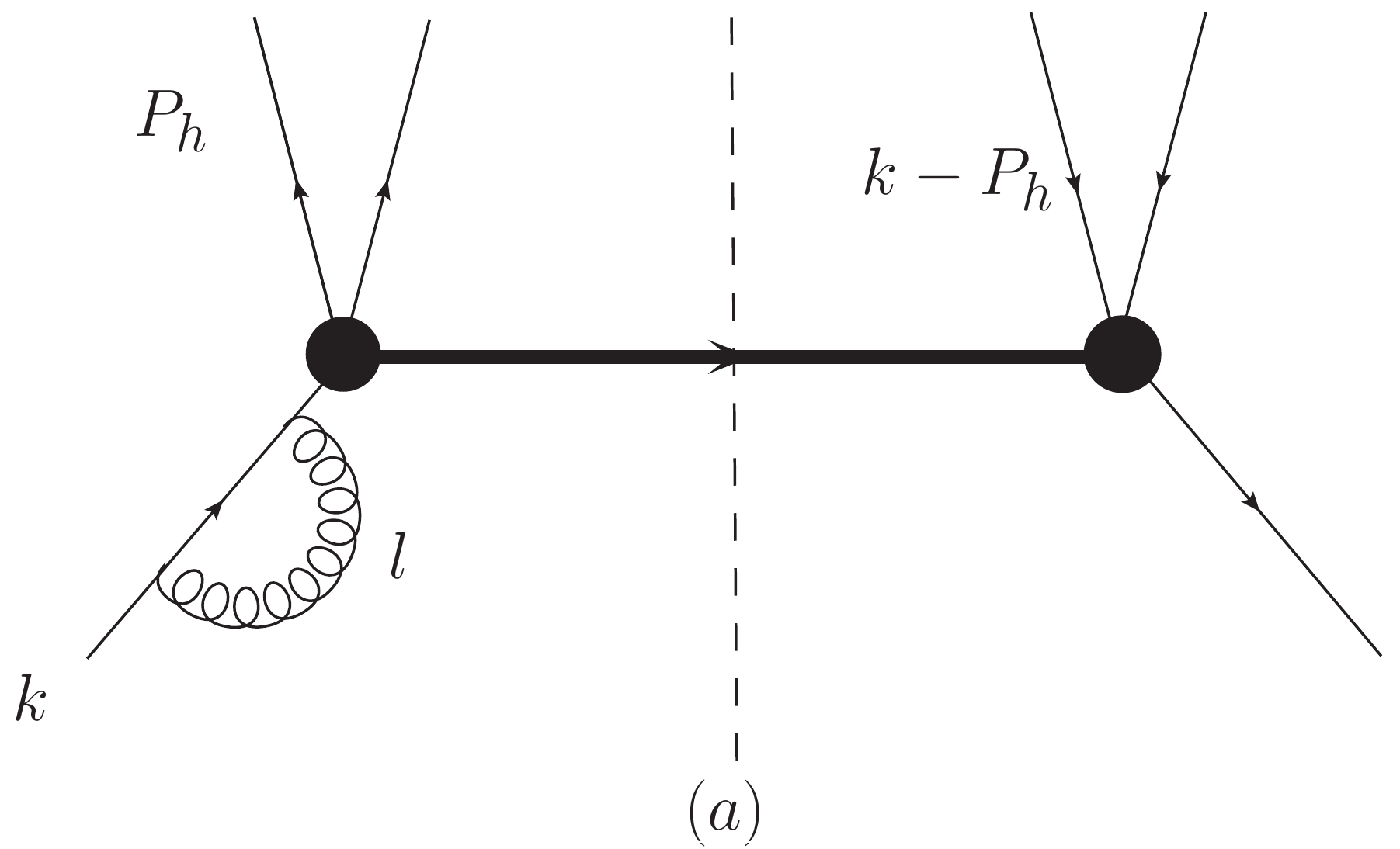}
\includegraphics[height=3cm,width=4cm]{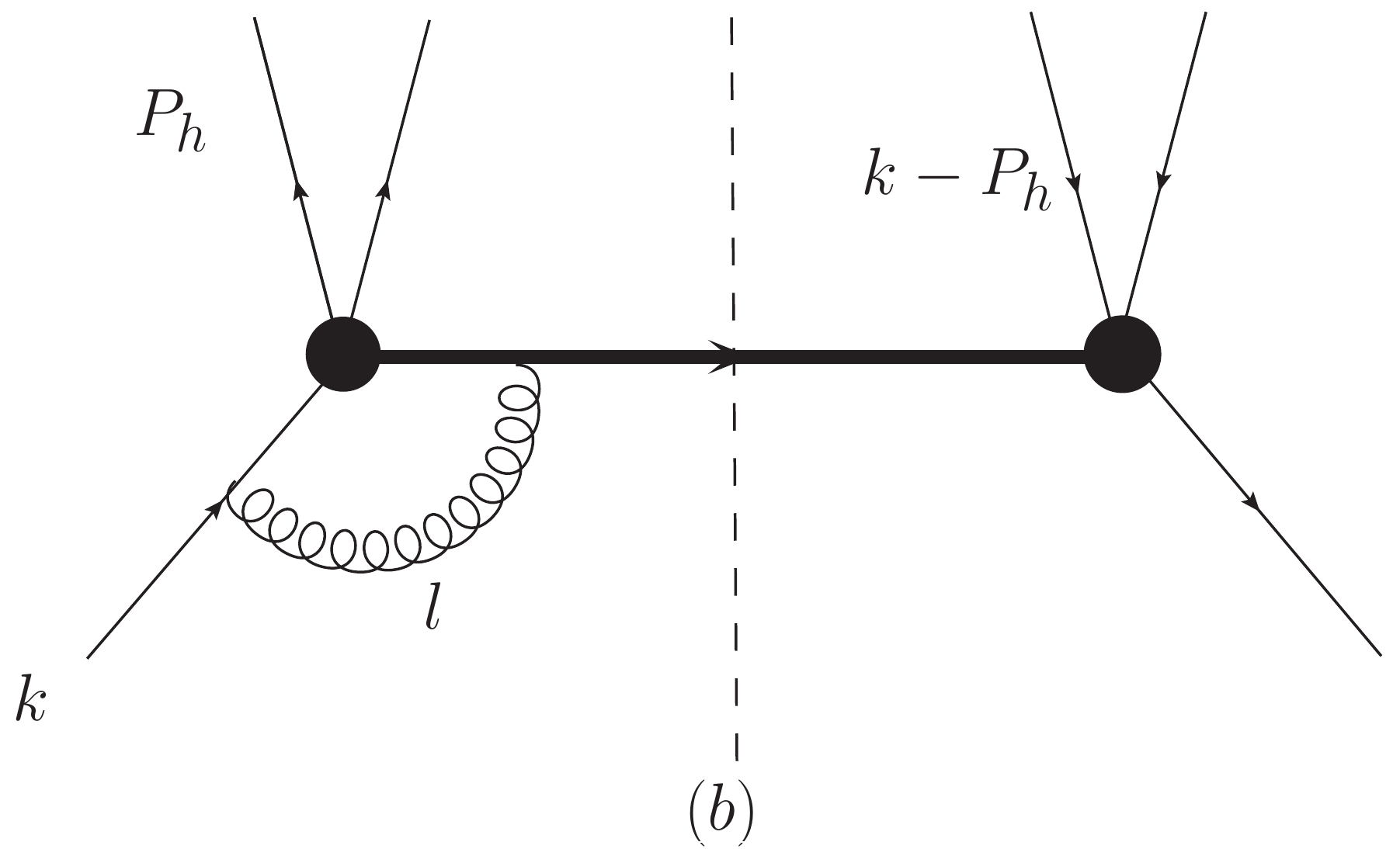}
\includegraphics[height=3cm,width=4cm]{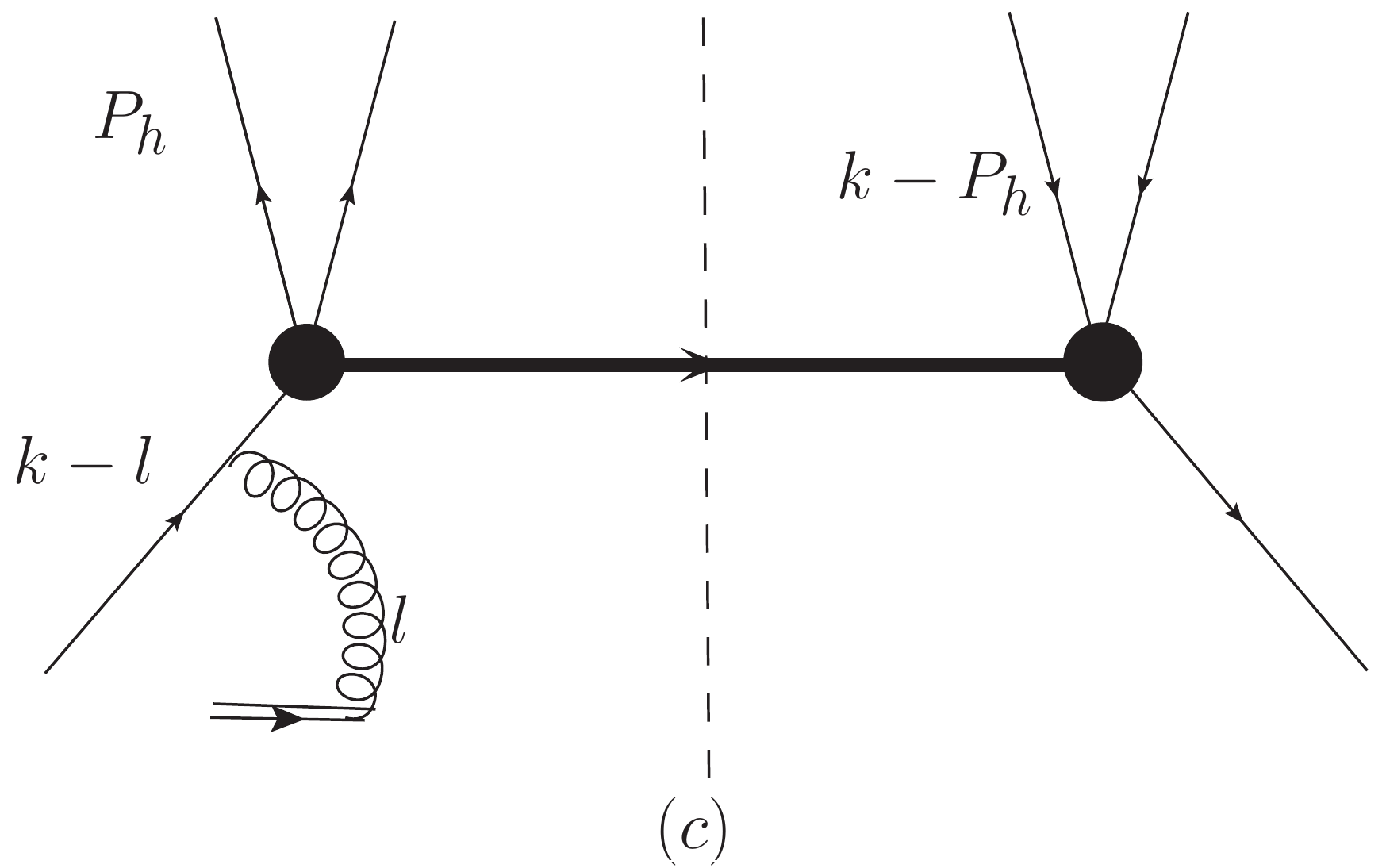}
\includegraphics[height=3cm,width=4cm]{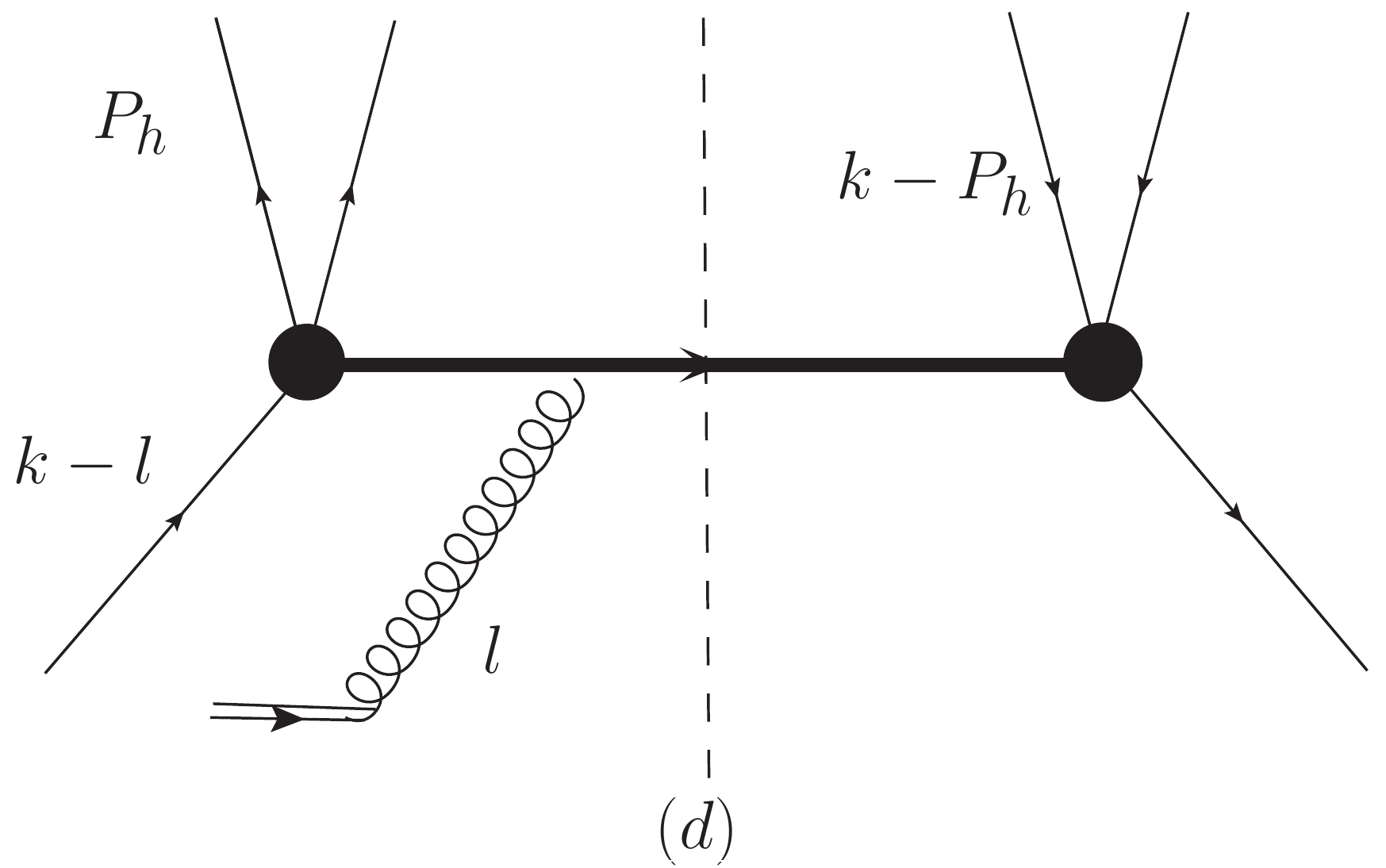}
+h.c.
\caption{ One loop order corrections to the fragmentation function of a quark into a meson pair in the spectator model. Where h.c. represents the hermitian conjugations of these diagrams.}
\label{fig:1}
\end{figure}
We can model the correlator at one loop level provided with Fig.\ref{fig:1} as:
\begin{eqnarray} \label{eq14} 
\begin{aligned}
\Delta_a^q(z,k_T^2,\cos\theta,M_h^2,\phi_R) &= i\frac{C_F \alpha_s}{32\pi^2 (1-z) P_h^-} \cdot \frac{ |\vec{R}|}{M_h} \cdot \frac{(\s{k}+m)}{(k^2-m^2)^3}\left( F^{s*}e^{-\frac{k^2}{\Lambda_s^2}}+F^{p*}e^{-\frac{k^2}{\Lambda_p^2}} \s{R} \right) (\s{k}-\s{P}_h+M_s) 
\\
&\left( F^{s}e^{-\frac{k^2}{\Lambda_s^2}}+F^{p}e^{-\frac{k^2}{\Lambda_p^2}} \s{R} \right) (\s{k}+m)
\int \frac{d^4 \ell}{(2\pi)^4} \frac{\gamma^\mu (\s{k}-\s{\ell}+m) \gamma_\mu (\s{k}+m)}{((k-\ell)^2-m^2+i\varepsilon)(\ell^2+i\varepsilon)} ,
\end{aligned}
\end{eqnarray}
\begin{eqnarray} \label{eq15}
\begin{aligned} 
\Delta_b^q(z,k_T^2,\cos\theta,M_h^2,\phi_R) &= i\frac{C_F \alpha_s}{32\pi^2 (1-z) P_h^-} \cdot \frac{ |\vec{R}|}{M_h} \cdot \frac{(\s{k}+m)}{(k^2-m^2)^2}\left( F^{s*}e^{-\frac{k^2}{\Lambda_s^2}}+F^{p*}e^{-\frac{k^2}{\Lambda_p^2}} \s{R} \right) (\s{k}-\s{P}_h+M_s) 
\\
&\int \frac{d^4 \ell}{(2\pi)^4} \frac{\gamma^\mu (\s{k}-\s{P}_h-\s{\ell}+M_s) \Big( F^{s}e^{-\frac{k^2}{\Lambda_s^2}}+F^{p}e^{-\frac{k^2}{\Lambda_p^2}} \s{R} \Big)  (\s{k}-\s{\ell}+m) \gamma_\mu (\s{k}+m)}{((k-P_h-\ell)^2-M_s^2+i\varepsilon)((k-\ell)^2-m^2+i\varepsilon)(\ell^2+i\varepsilon)} ,
\end{aligned}
\end{eqnarray}
\begin{eqnarray} \label{eq16} 
\begin{aligned}
\Delta_c^q(z,k_T^2,\cos\theta,M_h^2,\phi_R) &= i\frac{C_F \alpha_s}{32\pi^2 (1-z) P_h^-} \cdot \frac{ |\vec{R}|}{M_h} \cdot \frac{(\s{k}+m)}{(k^2-m^2)^2}\left( F^{s*}e^{-\frac{k^2}{\Lambda_s^2}}+F^{p*}e^{-\frac{k^2}{\Lambda_p^2}} \s{R} \right) (\s{k}-\s{P}_h+M_s) 
\\
&\left( F^{s}e^{-\frac{k^2}{\Lambda_s^2}}+F^{p}e^{-\frac{k^2}{\Lambda_p^2}} \s{R} \right) 
\int \frac{d^4 \ell}{(2\pi)^4} \frac{(\s{k}+m) \gamma^- (\s{k}-\s{\ell}+m) }{((k-\ell)^2-m^2+i\varepsilon)(-\ell^- \pm i\varepsilon)(\ell^2+i\varepsilon)} ,
\end{aligned}
\end{eqnarray}
\begin{eqnarray} \label{eq17} 
\begin{aligned}
\Delta_d^q(z,k_T^2,\cos\theta,M_h^2,\phi_R) &= i\frac{C_F \alpha_s}{32\pi^2 (1-z) P_h^-} \cdot \frac{ |\vec{R}|}{M_h} \cdot \frac{(\s{k}+m)}{k^2-m^2}\left( F^{s*}e^{-\frac{k^2}{\Lambda_s^2}}+F^{p*}e^{-\frac{k^2}{\Lambda_p^2}} \s{R} \right) (\s{k}-\s{P}_h+M_s) 
\\
&\int \frac{d^4 \ell}{(2\pi)^4} \frac{\gamma^- (\s{k}-\s{P}_h-\s{\ell}+M_s) \Big( F^{s}e^{-\frac{k^2}{\Lambda_s^2}}+F^{p}e^{-\frac{k^2}{\Lambda_p^2}} \s{R} \Big)  (\s{k}-\s{\ell}+m) }{((k-P_h-\ell)^2-M_s^2+i\varepsilon)((k-\ell)^2-m^2+i\varepsilon)(-\ell^- \pm i\varepsilon)(\ell^2+i\varepsilon)} .
\end{aligned}
\end{eqnarray}
The Feynman rule $1/(-\ell^- \pm i\varepsilon)$ have been applied for the eikonal propagator in Eq.(\ref{eq14}-\ref{eq17}) as well as that for the vertex between the eikonal line and the gluon.
Also in Eq.(\ref{eq14}-\ref{eq17}), in principle the Gaussian form factors should rely on the loop momentum $\ell$. Here resulting from the choice in Ref. \cite{Bacchetta:2007wc}, we get rid of this dependence and simply utilize $k^2$ rather than $(k-\ell)^2$ in those form factors to make straightforward the integration. This selection could given reasonable final results since the form factor is brought in to cut off the divergence. The same selection has also been assumed in Refs. \cite{Bacchetta:2002tk,Bacchetta:2003xn,Amrath:2005gv}. 

Since $G_{1,TT}^\perp$ comes from the interference of two p-waves, we have only one source in every diagrams at one loop level that is the imaginary part of the loop integral over $\ell$, coupling with the real quantity $|F^p|^2$. As for the imaginary part of the integral, we apply the Cutkosky cutting rules
\begin{eqnarray} \label{eq18}
\begin{aligned}
\frac{1}{\ell^2+i\varepsilon} \to -2\pi i \delta(\ell^2) \qquad \qquad \frac{1}{(k-\ell)^2+i\varepsilon} \to -2\pi i \delta((k-\ell)^2) .
\end{aligned}
\end{eqnarray}
Employing the above conventions, we reach the final result of $G_{1,TT}^\perp$
\begin{eqnarray} \label{eq19} 
\begin{aligned}
G_{1,TT}^{\perp a}&=0
\\
G_{1,TT}^{\perp b}&=\frac{1}{2\pi^3}\left[ \frac{C_F \alpha_s M_h |\vec{R}|^2}{(1-z)} \cdot |F^p|^2 e^{-\frac{2k^2}{\Lambda_p^2}} \right]\frac{1}{(k^2-m^2)^2} k_T C_b
\\
G_{1,TT}^{\perp c}&=0
\\
G_{1,TT}^{\perp d}&=-\frac{1}{2\pi^3}\left[ \frac{C_F \alpha_s M_h |\vec{R}|^2}{(1-z)} \cdot |F^p|^2 e^{-\frac{2k^2}{\Lambda_p^2}} \right]\frac{1}{k^2-m^2}( (I_2-\mathcal{A}) k_T )
\end{aligned}
\end{eqnarray}
with
\begin{eqnarray} \label{eq20}
\begin{aligned}
C_b &= (3k^2-m^2)\mathcal{A}+(k^2+2M_h^2-m^2+2mM_s-2M_s^2)\mathcal{B}
\\
&+(m^2-k^2)A_0+(m^2-M_h^2-2mM_s+M_s^2)B_0+(m^2-k^2)I_2 .
\end{aligned}
\end{eqnarray}
The coefficients $\mathcal{A}$ and $\mathcal{B}$ denote the following functions
\begin{eqnarray} \label{eq21}
\begin{aligned}
\mathcal{A} & =\frac{I_1}{\lambda(M_h,M_s)}\left[ 2k^2(k^2-M_s^2-M_h^2)\frac{I_2}{\pi}+(k^2+M_h^2-M_s^2) \right] \\
\mathcal{B} & =-\frac{2k^2}{\lambda(M_h^2,M_s^2)}I_1\left[ 1+\frac{k^2+M_s^2-M_h^2}{\pi}I_2 \right]
\end{aligned}
\end{eqnarray}
which originate from the decomposition of the following integral \cite{Lu:2015wja}
\begin{eqnarray} \label{eq22}
\begin{aligned}
\int d^4 \ell \frac{\ell^\mu \delta(\ell^2) \delta[(k-\ell)^2-m^2]}{(k-P_h-\ell)^2-M_s^2}=\mathcal{A}k^\mu + \mathcal{B}P_h^\mu .
\end{aligned}
\end{eqnarray}
The functions $I_i$ represent the results of the following integrals
\begin{eqnarray} \label{eq23}
\begin{aligned}
I_1&=\int d^4 \ell \delta(\ell^2) \delta[(k-\ell)^2-m^2]=\frac{\pi}{2k^2}(k^2-m^2) \\
I_2&=\int d^4 \ell \frac{\delta(\ell^2)\delta[(k-\ell)^2-m^2]}{(k-\ell-P_h)^2-M_s^2}=\frac{\pi}{2\sqrt{\lambda(M_h,M_s)}}\ln\left( 1-\frac{2\sqrt{\lambda(M_h,M_s)}}{k^2-M_h^2+M_s^2+\sqrt{\lambda(M_h,M_s)}} \right)
\end{aligned}
\end{eqnarray}
with $\lambda(M_h,M_s)=[k^2-(M_h+M_s)^2][k^2-(M_h-M_s)^2]$. 
In addition, the function $\mathcal{B}_0$ and $\mathcal{D}_0$ come from the decomposition
\begin{eqnarray} \label{eq24}
\begin{aligned}
\int d^4 \ell \frac{\ell^\mu \ell^\nu \delta(\ell^2) \delta((k-\ell)^2-m^2)}{(k-p-\ell)^2-M_s^2} &= k^\mu k^\nu \mathcal{A}_0+k^\mu p^\nu \mathcal{B}_0+p^\mu k^\nu \mathcal{C}_0+p^\mu p^\nu \mathcal{D}_0+g^{\mu\nu}\mathcal{E}_0 ,
\end{aligned}
\end{eqnarray}
where
\begin{eqnarray} \label{eq25}
\begin{aligned}
\mathcal{A}_0&=\frac{(k^2-m^2)(\mathcal{A}k^4-\mathcal{B}k^4-4\mathcal{A}k^2M_h^2-2\mathcal{B}k^2M_h^2+\mathcal{A}M_h^4+\mathcal{B}M_h^4-2\mathcal{A}k^2M_s^2+2\mathcal{B}k^2M_s^2-2\mathcal{A}M_h^2M_s^2+\mathcal{A}M_s^4-\mathcal{B}M_s^4)}{2k^2(k^4-2k^2M_h^2-2k^2M_s^2+M_h^4-2M_h^2M_s^2+M_s^4)}
\\
\mathcal{B}_0&=\frac{1}{2}\frac{(k^2-m^2)(\mathcal{A}k^2+3\mathcal{B}k^2+\mathcal{A}M_h^2-\mathcal{B}M_h^2-\mathcal{A}M_s^2-3\mathcal{B}M_s^2)}{k^4-2k^2M_h^2-2k^2M_s^2+M_h^4-2M_h^2M_s^2+M_s^4}.
\end{aligned}
\end{eqnarray}

\section{Numerical results}
\label{IV}

In order to fix the parameters of the spectator model, the authors of Ref.\cite{Bacchetta:2006un} compare it with the output of the PYTHIA event generator \cite{Sjostrand:2000wi} adopted for HERMES. The values of the parameters obtained by the fit are:
$
\alpha_s=2.60\ \text{GeV},\ \beta_s=-0.751,\ \gamma_s=-0.193,\ \alpha_p=7.07\ \text{GeV},\ \beta_p=-0.038,\ \gamma_p=-0.085,\ M_s=2.97M_h,\ f_s=1197\ \text{GeV}^{-1},\ f_\rho=93.5,\ f_\omega=0.63,\ f_\omega'=75.2
$.
For the quark mass $m$, we adopt the same choice as in Ref.\cite{Bacchetta:2006un} and fix it to be zero GeV. Notice that these model parameters are acquired by comparing the theoretical model with the PITHIA event generator adopted for the HERMES kinematics.
In the following we also make predictions in COMPASS and EIC kinematics, thus there exist uncertainties with regard to the model parameters.
In this paper, we make a rough consideration by disregarding such uncertainties. Furthermore, we choose the strong coupling $\alpha_s \approx 0.3$ for our preliminary estimation.

\begin{figure}[htp]
\centering
\includegraphics[height=5.0cm,width=6.7cm]{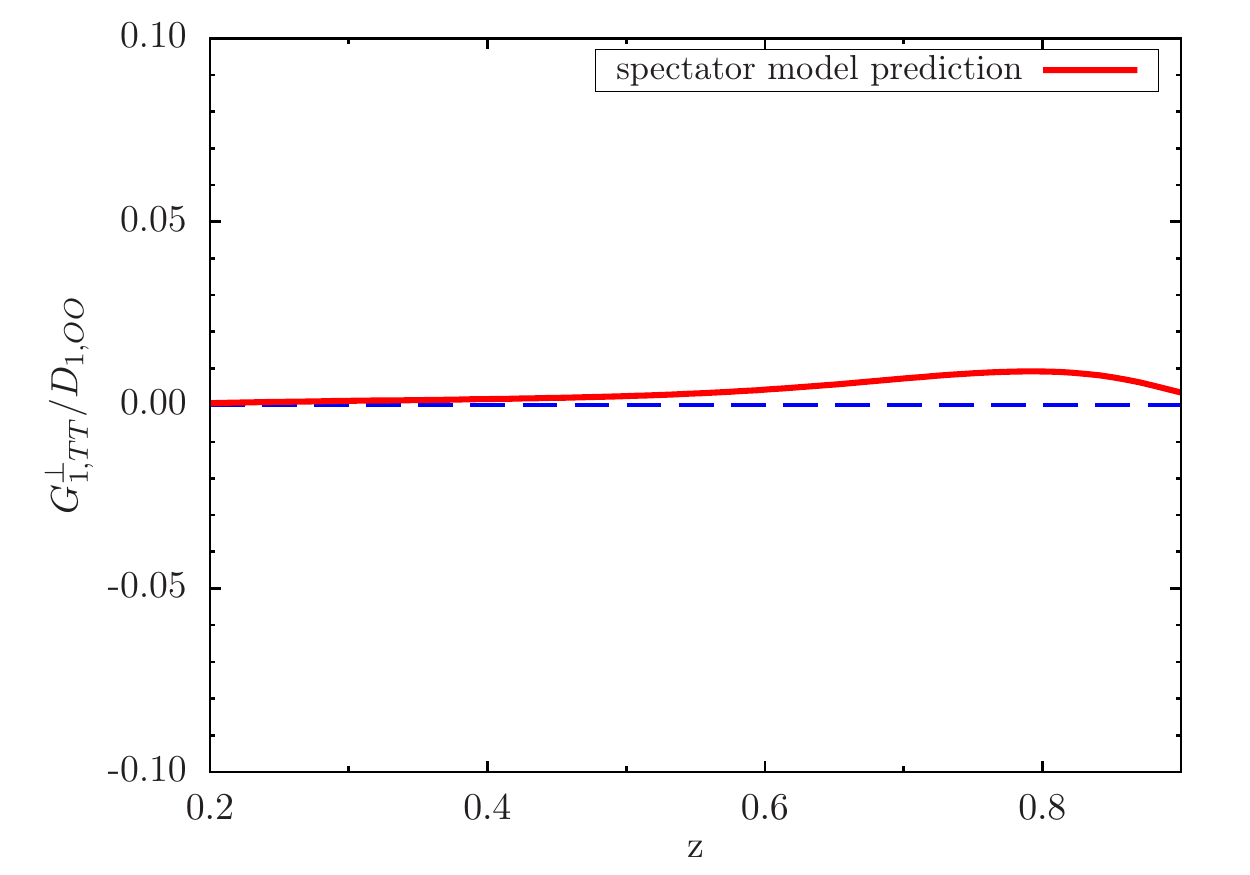}
\includegraphics[height=5.0cm,width=6.7cm]{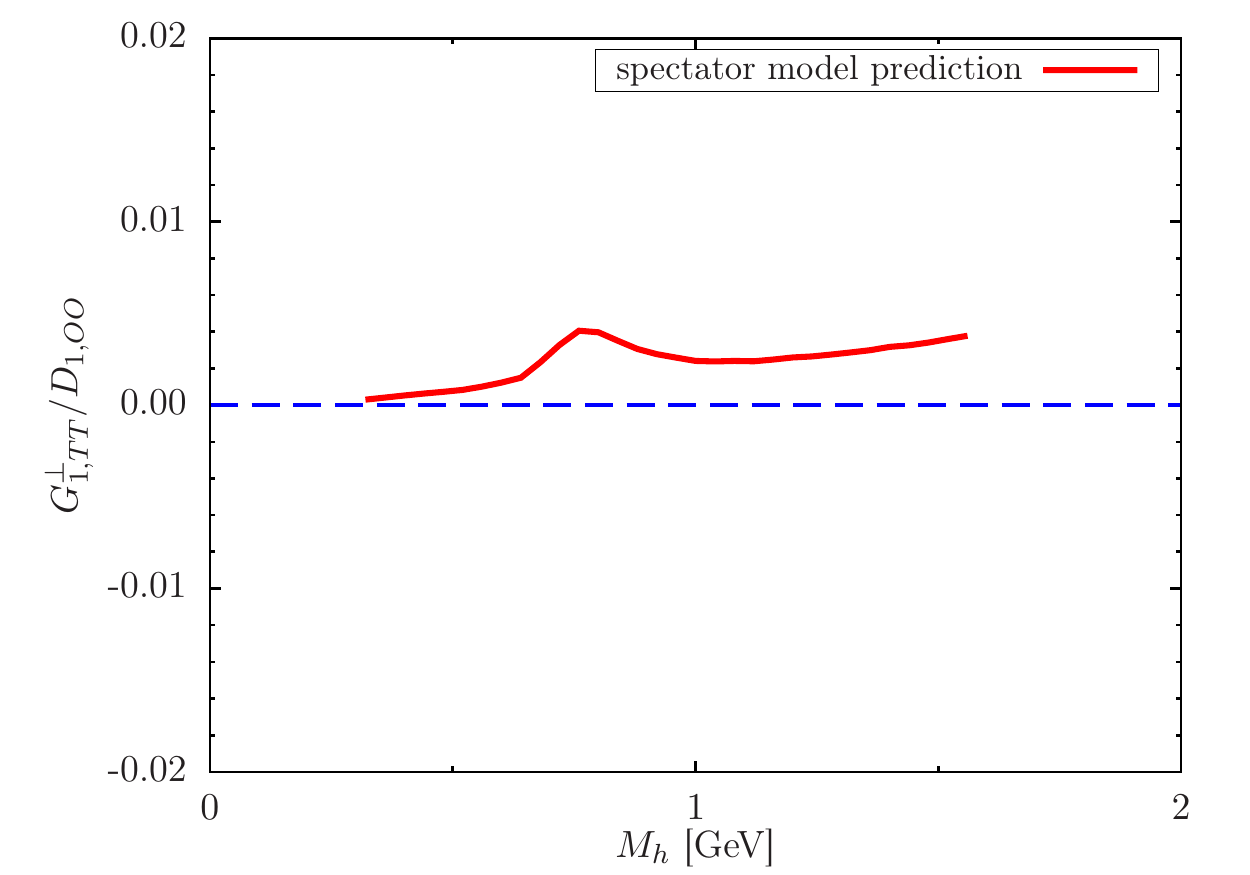}
\caption{\normalsize 
The DiFF $G^\perp_{1,TT}$ as functions of $z$ (left panel) and $M_h$ (right panel) in the spectator model, normalized by the unpolarized DiFF $D_{1,OO}$.}
\label{fig:2}
\end{figure}
Firstly, to quantify the magnitude of the DiFF $G^\perp_{1,TT}$, we plot the ratio between $G^\perp_{1,TT}$ and $D_{1,OO}$ as a function of $z$ or $M_h$, integrated over the region $0.3\ \text{GeV}<M_h<1.6\ \text{GeV}$ or $0.2<z<0.9$ in the left panel and right panel of Fig.\ref{fig:2} respectively. Here we have used the analytical result of the integrated $D_{1,OO}$ obtained in \cite{Bacchetta:2006un}. Comparing with the unpolarized DiFF $D_{1,OO}$, the $G^\perp_{1,TT}$ is three order of magnitude smaller and we can find a peak located nearly in $M_h=0.8$ GeV.

Then we present the study of the $\sin(2\phi_h-2\phi_R)$ azimuthal asymmetry in the SIDIS process with unpolarized muons scattering off longitudinally polarized nucleon target. Basing on the isospin symmetry, the fragmentation correlators for processes $u \to \pi^+ \pi^- X$, $\bar{d} \to \pi^+ \pi^- X$, $d \to \pi^- \pi^+ X$ and $\bar{u} \to \pi^- \pi^+ X$ are similar. Thus changing the sign of $\vec{R}$ equivalently implies replacing $\theta \to \pi-\theta$ by $\phi \to \phi+\pi$. While expanding the flavor sum in the numerator of Eq.(\ref{eq13}), $d \to \pi^- \pi^+ X$ and $\bar{u} \to \pi^- \pi^+ X$ contributions have the equal sign comparing to that of $u \to \pi^+ \pi^- X$ since the $\sin(2\phi_h-2\phi_R)$ modulation keeps its sign. Furthermore, in principle sea quark distributions can be produced via perturbative QCD evolution and they are zero at the model scale. In this paper, QCD evolution has been neglected, which leads to zero antiquark PDFs $f_1$ and $g_{1L}$. The expressions of the $x$-dependent, $z$-dependent and $M_h$-dependent $\sin(2\phi_h-2\phi_R)$ asymmetry can therefore be adopted from Eq.(\ref{eq13}) as follows
\begin{eqnarray} \label{eq26}
\begin{aligned}
&\displaystyle A_{UL}^{\sin(2\phi_h-2\phi_R)}(x) = \frac{\int\ {\cal N}_0\ dz\ dM_h\ d\cos\theta\ d^2\vec{P}_{h\perp} d^2 \vec{p}_T\ d^2 \vec{k}_T }{\int \ {\cal D}_0\ dz\ dM_h\ d\cos\theta\ d^2\vec{P}_{h\perp} d^2 \vec{p}_T\ d^2\vec{k}_T}
\\
&\displaystyle A_{UL}^{\sin(2\phi_h-2\phi_R)}(z) = \frac{\int {\cal N}_0\ dx\ dM_h\ d\cos\theta\ d^2\vec{P}_{h\perp} d^2 \vec{p}_T\ d^2 \vec{k}_T}{\int \ {\cal D}_0\ dx\ dM_h\ d\cos\theta\ d^2\vec{P}_{h\perp} d^2 \vec{p}_T\ d^2\vec{k}_T}
\\
&\displaystyle A_{UL}^{\sin(2\phi_h-2\phi_R)}(M_h) = \frac{\int\ {\cal N}_0\ dx\ dz\ d\cos\theta\ d^2\vec{P}_{h\perp} d^2 \vec{p}_T\ d^2 \vec{k}_T}{\int\ {\cal D}_0\ dx\ dz\ d\cos\theta\ d^2\vec{P}_{h\perp} d^2\vec{p}_T\ d^2\vec{k}_T}.
\end{aligned}
\end{eqnarray}
with
\begin{eqnarray} \label{eq27}
\begin{aligned}
{\cal N}_0 &= 2M_h \bigg[ 4g_{1L}^u(x,\vec{p}_T^2)+g_{1L}^d(x,\vec{p}_T^2) \bigg] \sin\theta \ \delta\left( \vec{p}_T-\vec{k}_T-\frac{\vec{P}_{h\perp}}{z}  \right) \frac{2(\vec{k}_T \cdot \hat{P}_{h\perp})^2-\vec{k}_T^2}{M_h^2} \left( \frac{|\vec{R}|}{2|\vec{k}_T|} G_{1,TT}^\perp(z,\vec{k}_T^2,M_h) \right),
\\
{\cal D}_0&= 2M_h \bigg[4f_{1}^u(x,\vec{p}_T^2)+f_{1}^d(x,\vec{p}_T^2) \bigg] \ \delta\left( \vec{p}_T-\vec{k}_T-\frac{\vec{P}_{h\perp}}{z}  \right) D_{1,OO}(z,\vec{k}_T^2,M_h).
\end{aligned}
\end{eqnarray}
Here for the twist-2 PDFs $f_1$ and $g_1$, we apply the same spectator model results \cite{Bacchetta:2008af} for uniformity.   
The TMD DiFF $D_{1,OO}(z,\vec{k}_T^2,M_h^2)$ has been worked out and listed as \cite{Luo:2019frz}
\begin{eqnarray} \label{eq28}
\begin{aligned}
D_{1,OO}(z,\vec{k}_T^2,M_h)=&\frac{4\pi|\vec{R}|}{256\pi^3M_h z (1-z)(k^2-m^2)^2}\bigg\{ 4|F^s|^2 e^{-\frac{2k^2}{\Lambda_s^2}} (zk^2-M_h^2-m^2z+m^2+2mM_s+M_s^2)
\\
&-4|F^p|^2 e^{-\frac{2k^2}{\Lambda_p^2}}|\vec{R}|^2 (-zk^2+M_h^2+m^2(z-1)+2mM_s-M_s^2)
\\
&+\frac{4}{3} |F^p|^2 e^{-\frac{2k^2}{\Lambda_p^2}} |\vec{R}|^2 \bigg[4\left( \frac{M_h}{2z}-z\frac{k^2+\vec{k}_T^2}{2M_h} \right)^2 +2z \frac{k^2-m^2}{M_h} \left( \frac{M_h}{2z}-z\frac{k^2+\vec{k}_T^2}{2M_h} \right) \bigg] \bigg\}.
\end{aligned}
\end{eqnarray}

To perform numerical calculation for the $\sin(2\phi_h-2\phi_R)$ asymmetry in dihadron SIDIS, we adopt the kinematical cuts at the COMPASS, HERMES and EIC measurements as
\begin{itemize} \label{eq29}
\item \text{Cut1} \cite{Sirtl} at the COMPASS{\color{red}}: $\sqrt{s} = 17.4\ \text{GeV}, ~ 0.003 < x < 0.4, ~  0.1 < y < 0.9, ~ 0.2 < z < 0.9, ~ 0.3\text{GeV} < M_h < 1.6\ \text{GeV}, ~ Q^2 > 1\ \text{GeV}^2, ~ W > 5\ \text{GeV}$,
\item \text{Cut2} \cite{Airapetian:2009ae} at the HERMES {\color{red}}: $\sqrt{s} = 7.2\ \text{GeV}, ~ 0.023 < x < 0.4, ~ 0.1 < y < 0.95, ~ 0.2 < z < 0.7, ~ 0.3\ \text{GeV} < M_h < 1.6\ \text{GeV}, ~ Q^2 > 1\ \text{GeV}^2, ~ W^2 > 10\ \text{GeV}^2$,
\item \text{Cut3} \cite{Accardi:2012qut} at the EIC {\color{red}}: $\sqrt{s} = 45\ \text{GeV}, ~ 0.001 < x < 0.4, ~ 0.01 < y < 0.95, ~ 0.2 < z < 0.8, ~ 
0.3\ \text{GeV} < M_h < 1.6\ \text{GeV}, ~ Q^2 > 1\ \text{GeV}^2, ~ W^2 > 10\ \text{GeV}^2$,
\end{itemize}
respectively, where $W$ is the invariant mass of photon-nucleon system with $W^2=(P+q)^2 \approx \frac{1-x}{x}Q^2$.

\begin{figure}[htp]
\centering
\includegraphics[height=5.0cm,width=5.7cm]{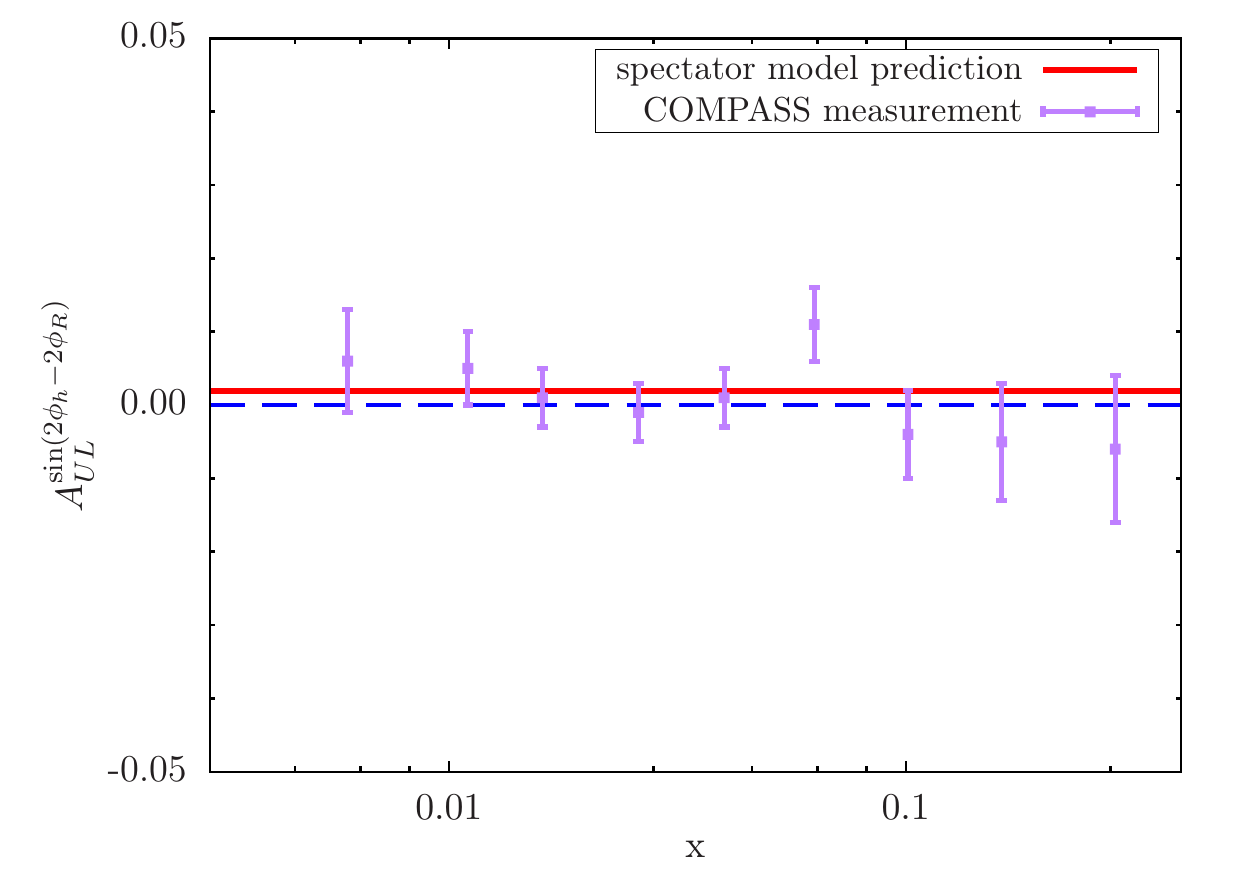}
\includegraphics[height=5.0cm,width=5.7cm]{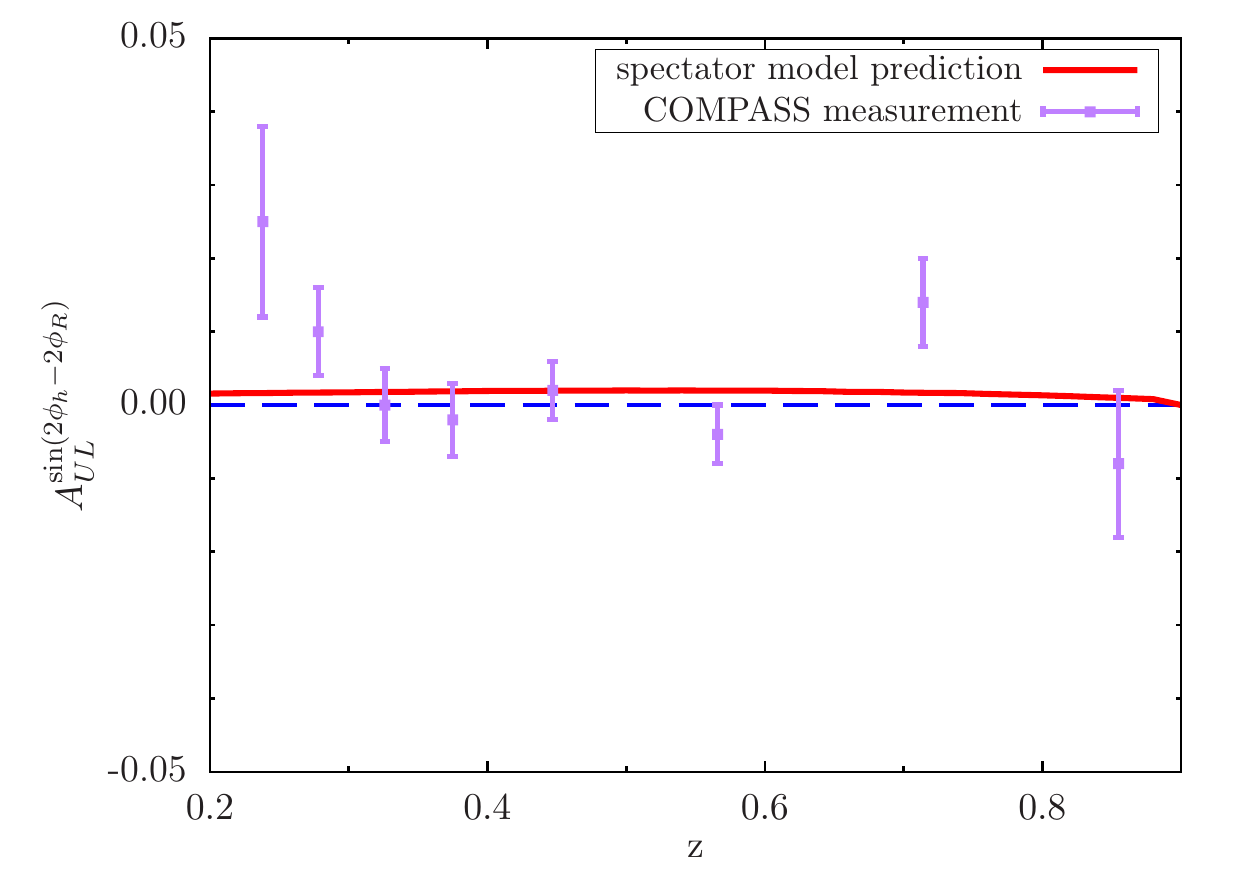}
\includegraphics[height=5.0cm,width=5.7cm]{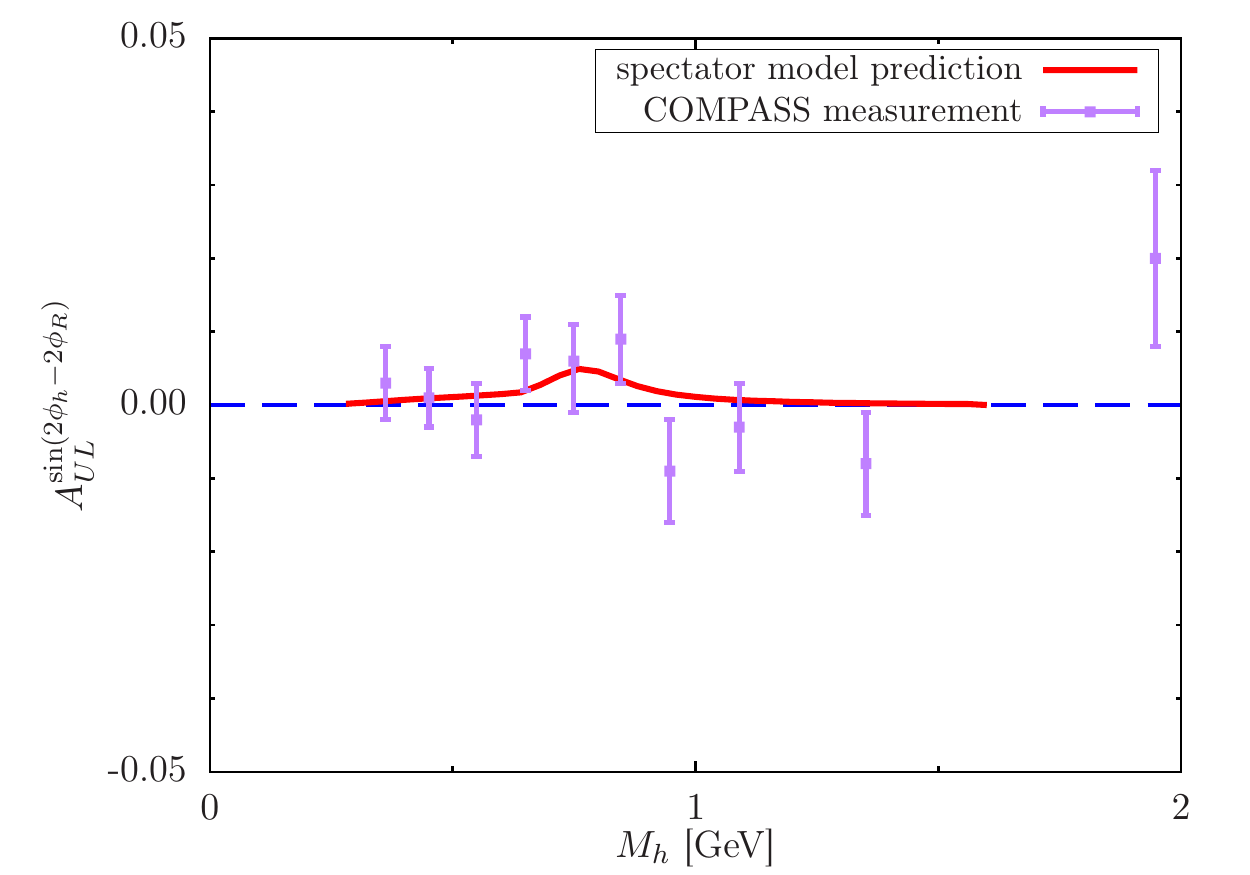}
\caption{\normalsize 
The $\sin(2\phi_h-2\phi_R)$ azimuthal asymmetry in the SIDIS process of unpolarized muons off longitudinally polarized nucleon target as a functions of $x$ (left panel), $z$ (central panel) and $M_h$ (right panel) at COMPASS. The full circles with error bars show the preliminary COMPASS data for comparison. The solid curves denote the model prediction.}
\label{fig:3}
\end{figure}

Our main results are plot in Fig.\ref{fig:3}, showing the predictions for the $\sin(2\phi_h-2\phi_R)$ azimuthal asymmetry. The $x$-, $z$- and $M_h$-dependent asymmetries are depicted in the left, central and right panels of the figure, respectively. The solid lines represent our model predictions. The full circles with error bars show the preliminary COMPASS data for comparison. We can find that the model predictions give a good description of the COMPASS preliminary data being compariable with zero. The model predicts a small peak in $M_h$-distribution. According to result of $G_{1,TT}^\perp$ based on the model calculation, the small DiFF may be one of the reason that result in such small asymmetry.
\begin{figure}[htp]
\centering
\includegraphics[height=5cm,width=5.7cm]{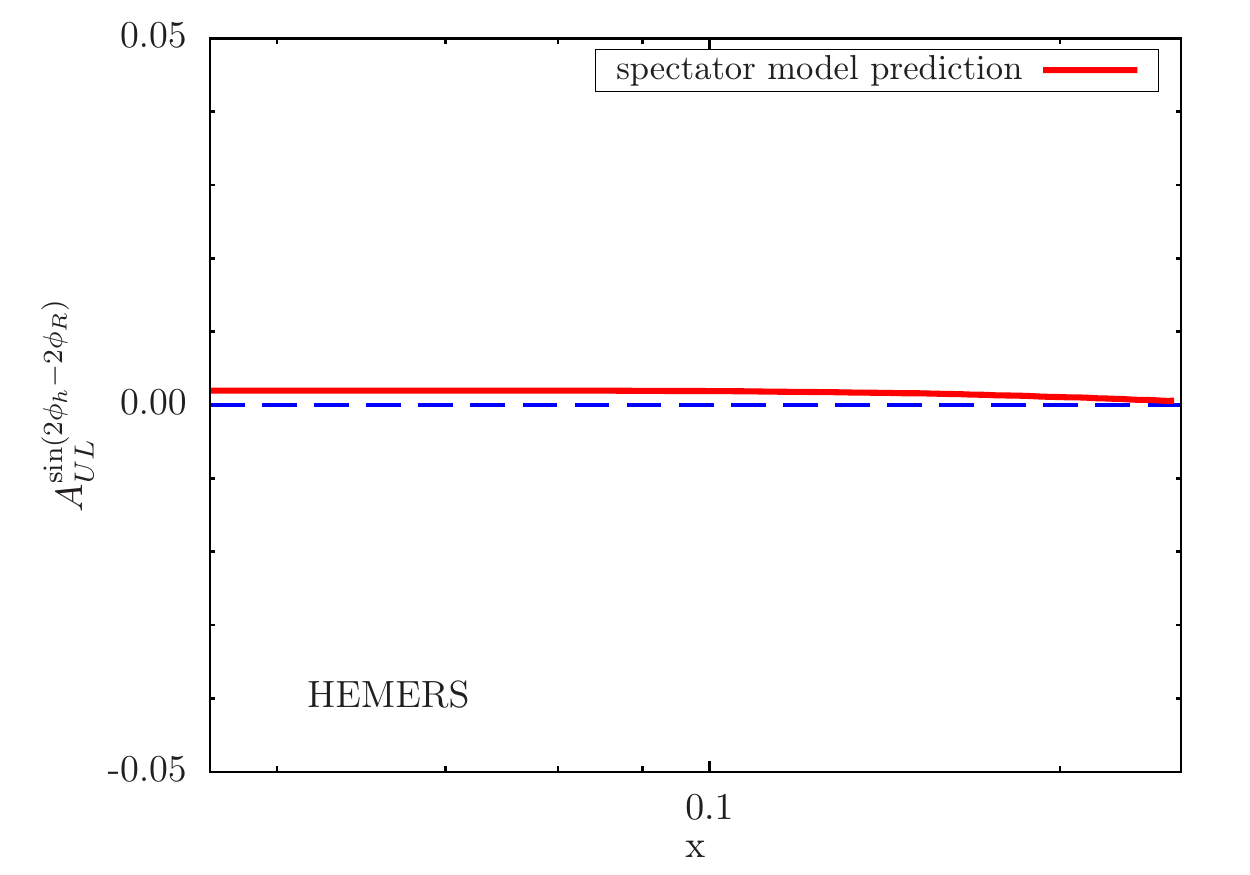}
\includegraphics[height=5cm,width=5.7cm]{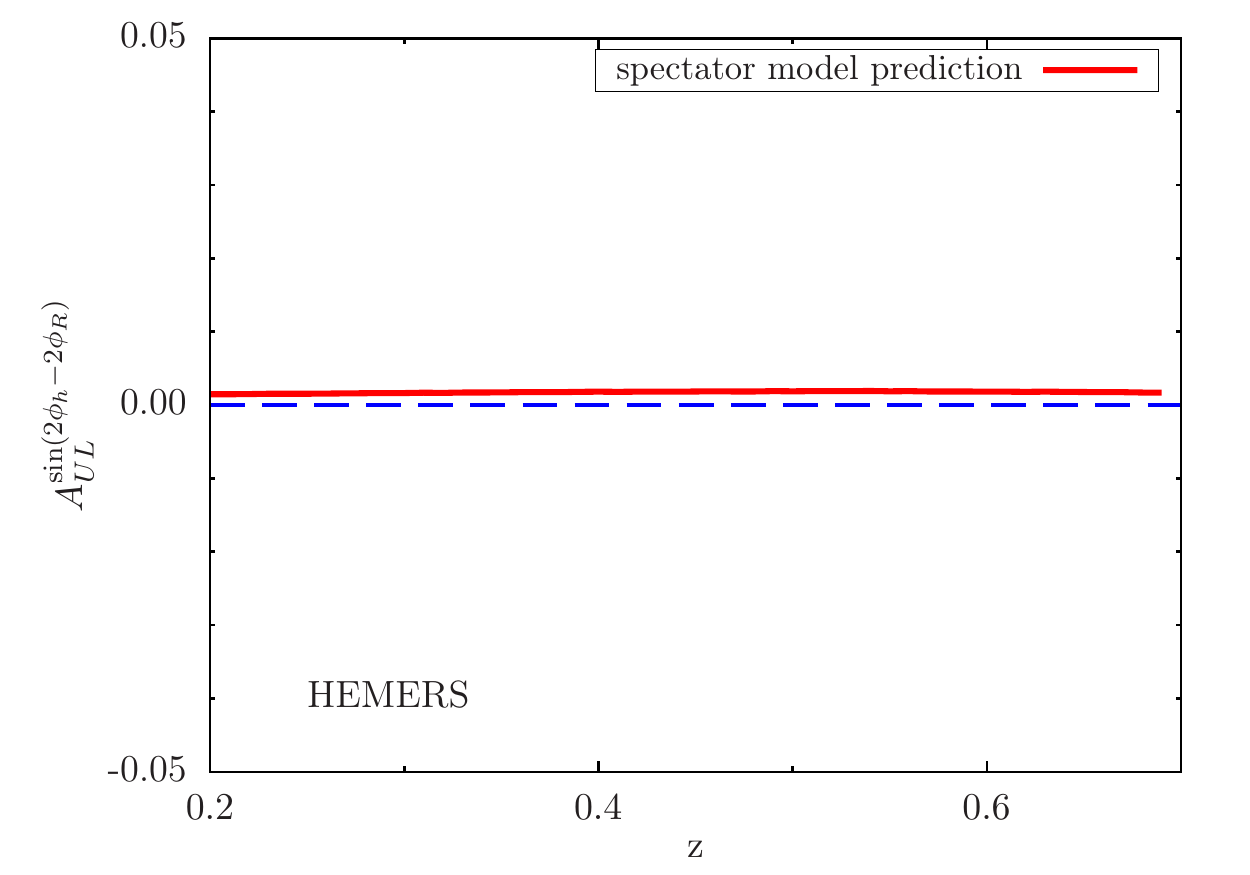}
\includegraphics[height=5cm,width=5.7cm]{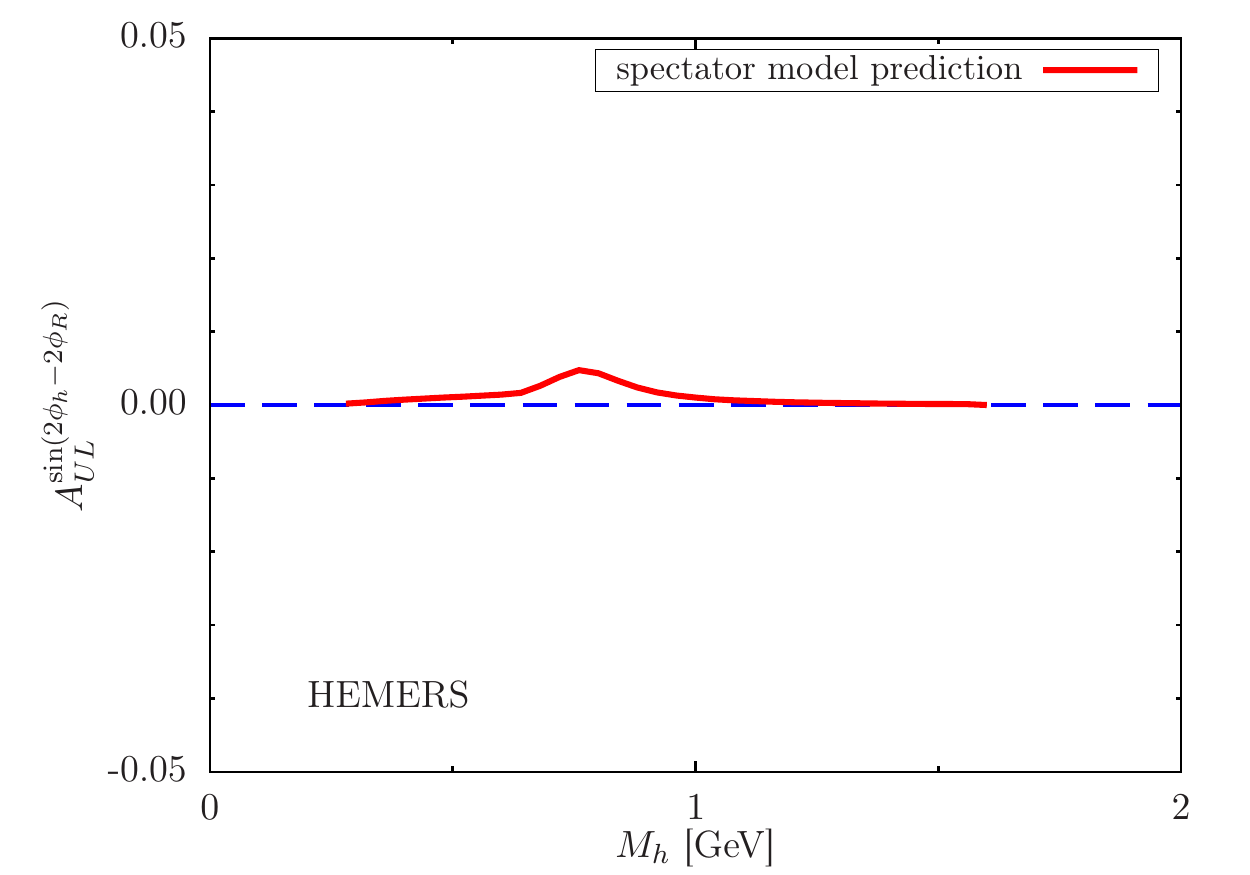}
\includegraphics[height=5cm,width=5.7cm]{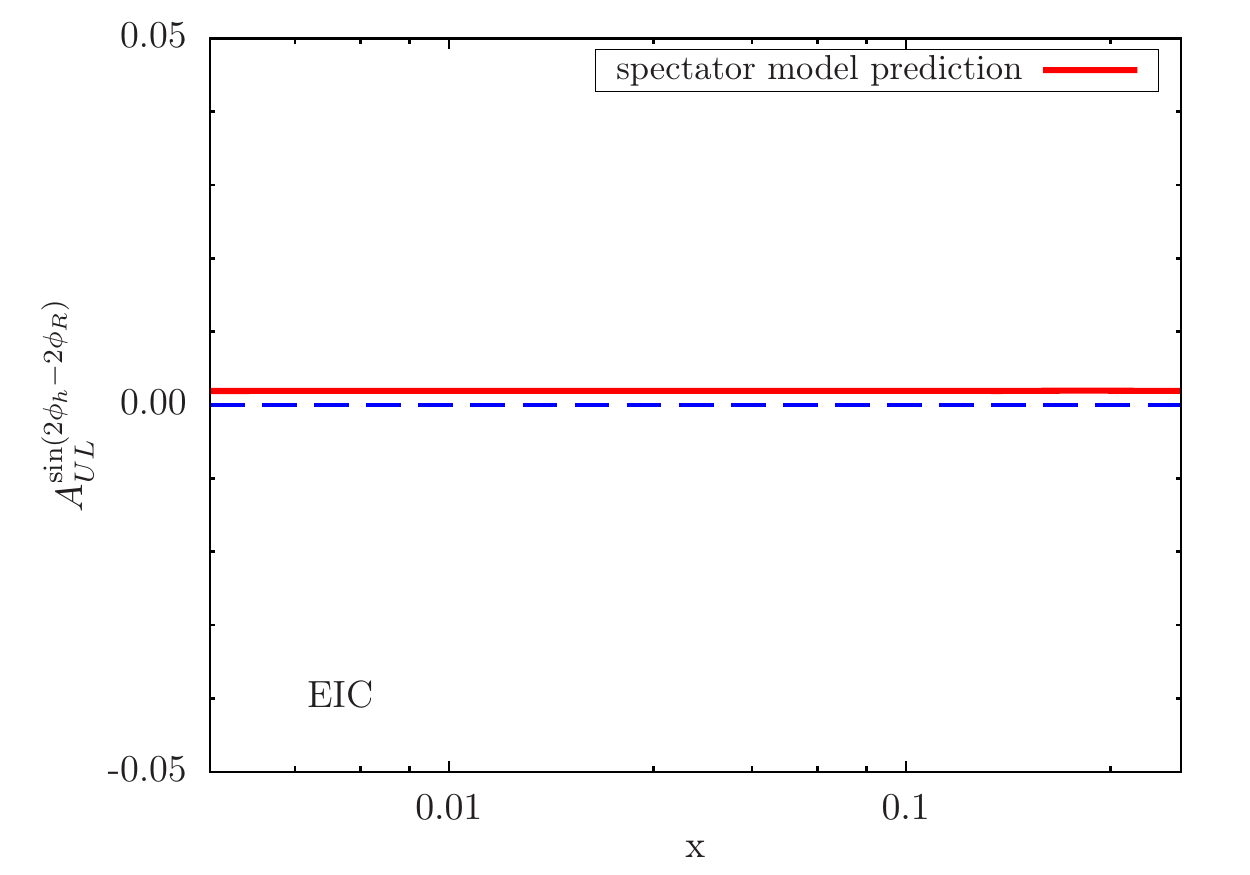}
\includegraphics[height=5cm,width=5.7cm]{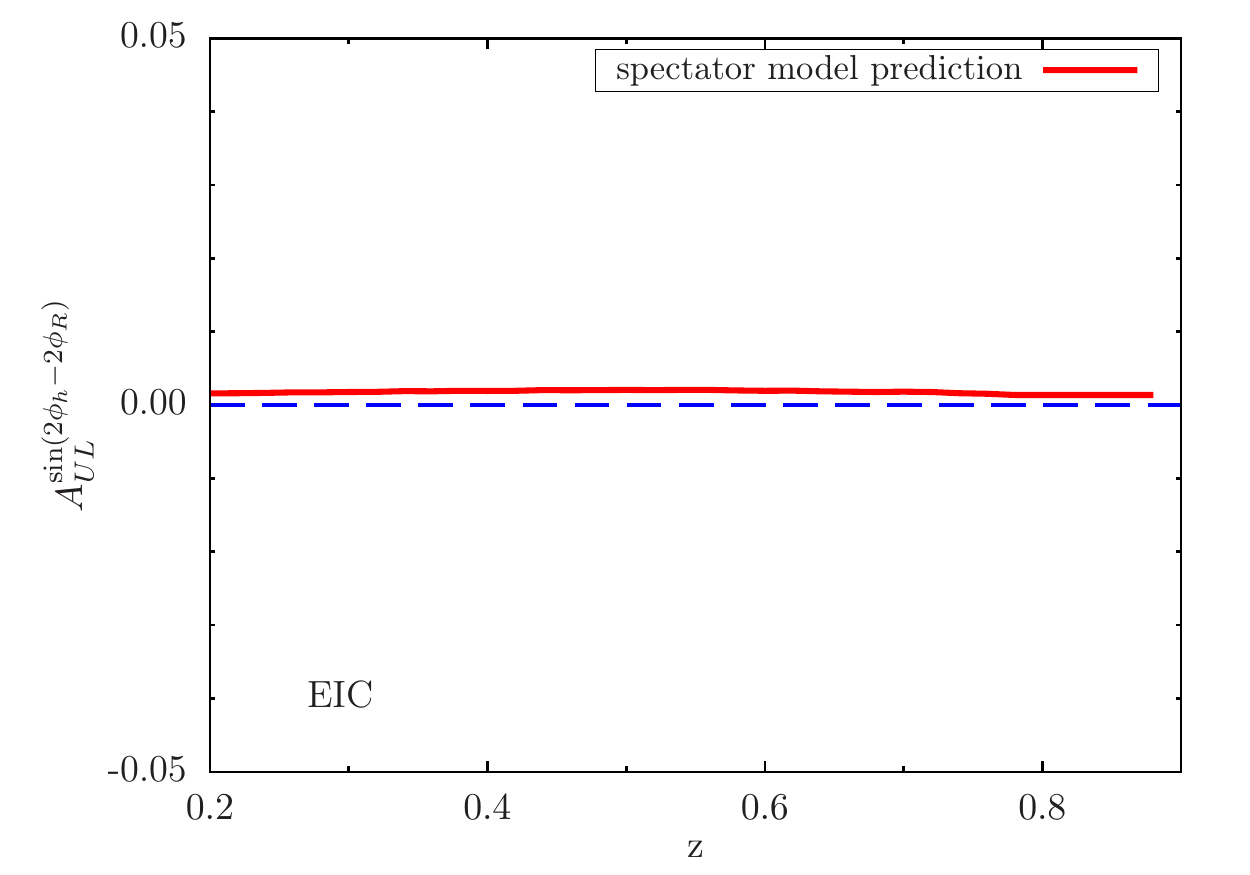}
\includegraphics[height=5cm,width=5.7cm]{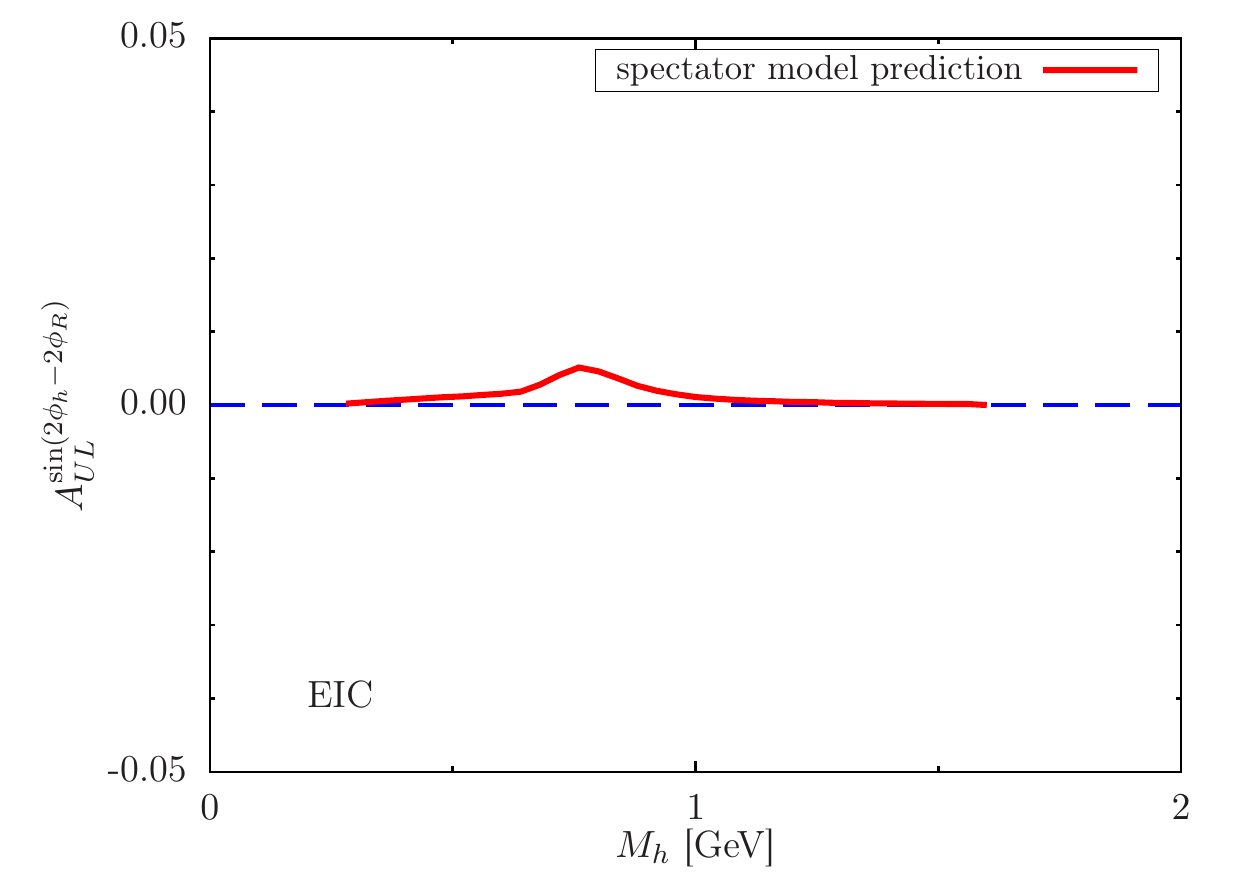}
\caption{\normalsize
The $\sin(2\phi_h-2\phi_R)$ azimuthal asymmetry in the SIDIS process 
of unpolarized muons off longitudinally polarized nucleon target as a functions of $x$ (left panel), $z$ (central panel) 
and $M_h$ (right panel) at the HERMES. The three lower panels display the same azimuthal asymmetry but at the EIC. The solid curves denote the model prediction. }
\label{fig:4}
\end{figure}
In order to make a further comparison, we also obtain the $\sin(2\phi_h-2\phi_R)$ asymmetry at the HERMES with kinematical Cut2 \cite{Airapetian:2009ae} and EIC with Cut3 \cite{Accardi:2012qut}.
The $x$-, $z$- and $M_h$-dependent asymmetries are plotted in the upper left, central, and right panels and the lower left, central, and right panels in Fig.\ref{fig:4}, respectively. 
We find that the overall tendency of the asymmetry at both the HERMES and EIC are similar to that at COMPASS. 
The size of the asymmetries are slightly smaller than that at COMPASS, and are still compariable with zero at the kinematics of HERMES and EIC. Similar predictions can also be achieved at the CLAS12 \cite{Matevosyan:2015gwa} and at the EicC{\color{red}} \cite{Chen:2019equ}.

\section{Conclusion}\label{V}

The single spin asymmetry with a $\sin(2\phi_h-2\phi_R)$ modulation of dihadron production in SIDIS is studied in this work. 
The T-odd DiFF $G_{1,TT}^\perp$ by taking the real and imaginary loop contributions is worked out with the accessible spectator model result for $D_{1,OO}$. $G_{1,TT}^\perp$ is originated from the interference contribution of two $p$-waves with the use of partial wave expansion. 
We find that one must consider loop contributions to obtain a nonvanishing $G_{1,TT}^\perp$. The prediction for $\sin(2\phi_h-2\phi_R)$ asymmetry is presented and compared with the COMPASS measurement by the means of the numerical results of the DiFFs and PDFs. Our result yields a good description of the vanished COMPASS data. At the HERMES and EIC kinematics we also obtain a very small asymmetry.

\begin{acknowledgments}
Xuan Luo thanks professors Marco Radici and Alessandro Bacchetta for their patient guidance. 
Hao Sun is supported by the National Natural Science Foundation of China (Grant No.11675033) 
and by the Fundamental Research Funds for the Central Universities (Grant No. DUT18LK27).
\end{acknowledgments}
\bibliography{v1}
\end{document}